\documentstyle[aas2pp4]{article}

\def\,{\thinspace}

\def\kms{km\thinspace s$^{-1}$}
\def\Kkms{K\thinspace km\thinspace s$^{-1}$}

\def\Msun{M$_\odot$}

\received{1996 November}
\accepted{}
\journalid{337}{15 January 1989}
\articleid{11}{14}
\slugcomment{}

\begin{document}
\title{Neutral atomic carbon in the globules of the Helix}

\author{K. Young}
\affil{Smithsonian Astrophysical Observatory,
60 Garden Street MS 78, Cambridge, MA 02138}
\author{P. Cox}
\affil{Institut d'Astrophysique Spatiale, Universit\'e de Paris XI,
91405 Orsay, France}
\affil{Institut d'Astrophysique de Paris, 92 bis, Boulevard Arago, 
75014 Paris, France}
\author{P.J. Huggins}
\affil{Physics Department, New York University, 4 Washington Place,
New York, NY 10003}
\author{T. Forveille}
\affil{Observatoire de Grenoble, B.P. 53X, 38041 Grenoble Cedex, France}
\author{R. Bachiller}
\affil{IGN Observatorio Astronomico Nacional, Apt 1143, 28800 Alcal\`{a} 
de Henares, Spain}

\begin{abstract}
We report detection of the 609~$\rm \mu m$ 
$\rm ^3P_1 \rightarrow {^3P_0}$ line of neutral atomic carbon in 
globules of the Helix nebula. The measurements were made
towards the position of peak CO emission. At the same position, we
obtained high-quality CO(2--1) and $^{13}$CO(2--1) spectra
and a 135$\arcsec \times 135 \arcsec$ map in CO(2--1).
The velocity distribution of C\,I shows six narrow 
(1--2~km~s$^{-1}$) components which are associated with individual 
globules traced in CO. The C\,I column densities are 
0.5--1.2$\times$10$^{16}$\,cm$^{-2}$. C\,I is found to be 
a factor of $\sim$6 more abundant than CO.
The large abundance of C\,I in the Helix can 
be understood as a result of the gradual photoionisation
of the molecular envelope
by the central star's radiation field.
\end{abstract}


\section{Introduction}
The Helix (NGC~7293, PK~36--57.1) is the nearest planetary
nebula (PN) with a massive molecular envelope
(Huggins et al.\ 1996).  At a distance of
$\sim$160~pc (e.g., Cahn et al.\ 1992) it affords the best
opportunity to explore in detail the relation between neutral
and ionized gas in an evolved nebula.  
Molecular gas in the Helix forms a large
broken ring surrounding the ionized cavity (Healy \& Huggins
1986; Cox et al., in preparation),
and its mass is at least
0.03 \Msun, or $\gtrsim25$\% of the total nebular
mass (Huggins \& Healy 1986).  High-resolution observations show that
the molecular gas is fragmented into numerous substructures
(Forveille \& Huggins 1991), some of which survive within the
ionized cavity and are seen as the well known cometary globules
(Huggins et al.\ 1992). HST studies of these remarkable structures
have been reported by O'Dell \& Handron (1996).

The molecular gas in PN is exposed to strong UV radiation
from the central star which is expected to establish
photodissociation regions (PDRs) with an important component of atomic
gas at the interface with the ionized
gas. A characteristic of such regions 
is emission in the ground state $\rm ^3P_1 \rightarrow {^3P_0}$
fine structure line of neutral atomic carbon at 609~$\rm\mu m$.  
This line has been detected in the young PN NGC~7027 (Young et al.,
in preparation) where C\,I is found to be a
minor constituent. It has also been
detected in the Ring nebula (Bachiller et al.\
1994) where the abundance of C\,I exceeds that of CO, although the
detailed structure of the gas is not resolved.

In this letter we report the detection of the 609~$\rm \mu m$ C\,I
line in the Helix. The observations, made 
towards the position of peak CO emission, reveal multiple, narrow (1--2 {\kms})
C\,I components which we unambigouly associate with the globules
through mapping the components in CO. 
The observations show that CI is a major constituent of the 
neutral gas in the globules.

\section{Observations and results}

The observations were made using the 10.4~m telescope of the Caltech
Submillimeter Observatory (CSO)\thanks{The Caltech Submillimeter
Observatory is funded by the National Science Foundation under
contract AST-9015755}.
The region observed lies on the western limb of the ionized nebula and
corresponds to the position of peak CO emission (see Fig.~1).
The observations include deep
integrations at the field center in  C\,I, CO(2--1), and  
$^{13}$CO(2--1), and mapping 
of the region in CO(2--1) to determine the
local structure of the~gas.

\begin{figure}[h]
\plotone{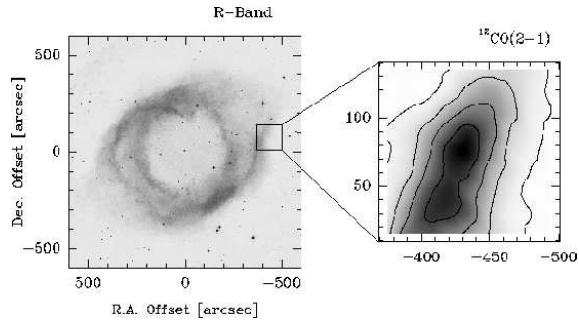}
\caption{Location of the region observed in C~I
and in CO shown by the inset on 
an optical image (R-band) of the Helix nebula. The velocity integrated
$^{12}$CO(2$-$1) map shows the distribution of the molecular gas in this 
region. Contours are 1 to 12 by 3 K~km~s$^{-1}$.  The field center is
offset by ($-$435$\arcsec$,
75$\arcsec$) from the central star, whose coordinates are
$22^{\rm h}26^{\rm m}54.7^{\rm s}$, $-21\arcdeg 05\arcmin 
37\arcsec$ (1950.0).} \label{onebarrel}
\end{figure}

The C\,I observations at 609~$\rm \mu m$ (492 GHz) were obtained in July 1996
in excellent weather conditions.
 The SSB system temperature (${\rm T_{SSB}}$) of the receiver was typically
$\sim 1900$K, and the spectrometer's resolution was 1 km s$^{-1}$.
The beam size  at
492~GHz was $15 \arcsec$ (FWHM).  We checked pointing by 
mapping CO(4--3) in the nearby
star EP Aqr and found its error to be  $\leq 4\arcsec$.
A chopping secondary mirror was used to observe 2 reference positions
$\pm 500\arcsec$ from the source in azimuth.
Observations were taken only when the source was
near transit to ensure that the reference positions were
locations previously found to be free of CO emission.

We also obtained 
CO(2--1) and $^{13}$CO(2--1) spectra at the 
position observed in C\,I
and mapped a region of 135$\arcsec \times 135 \arcsec$ around
this position in CO(2--1) with 15$\arcsec$ spacing. 
 For these
observations ${\rm T_{SSB}}$ was $\sim 250$ K,
and the spectrometer's resolution was $\sim$0.1~km~s$^{-1}$.  All
intensities in this letter are in units of main beam temperature.

The C\,I, CO and $^{13}$CO  spectra are shown in Fig.~2.
The three spectra are characterized by multiple velocity components,
most clearly seen in the CO spectrum. There are at least five 
separate components at velocities 
between $-$28 and $-$9 {\kms}, and additional, weaker components 
around $-$30 \kms\ and possibly $-$6 \kms. 
The strongest
components can be identified in all three spectra.

\begin{figure}[h]
\plotone{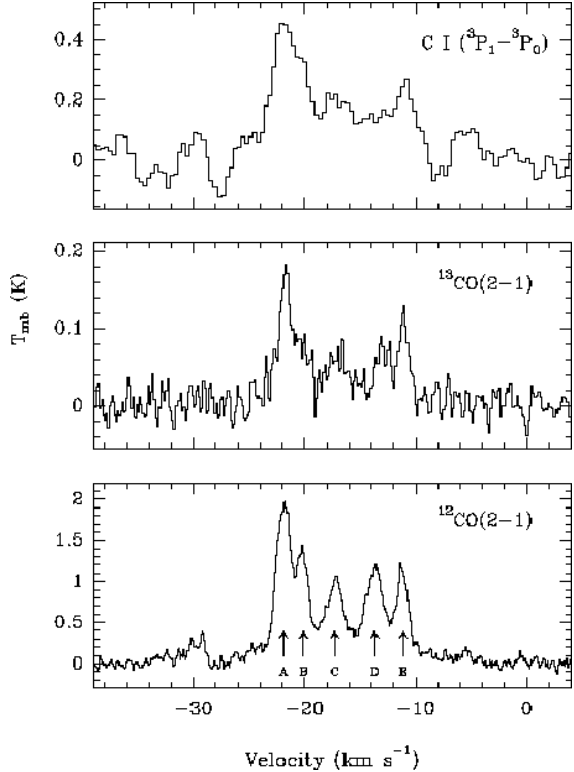}
\caption{
C\,I, $^{13}$CO(2$-$1) and CO(2$-$1) spectra towards the peak CO
position on the western limb of the Helix. The components
A to E (see text) are indicated in the lower panel.
The position is offset by
($-$435$^{\prime\prime}$, 75$^{\prime\prime}$) from 
the central star.
} \label{twobarrel}
\end{figure}

To characterize the spectra, we made multi-Gaussian fits to
the data. For the CO spectrum, we fit the five main
components with the velocities ($v_o$), intensities ($T$), and
linewidths ($\Delta v$) as free parameters. These values for $v_o$ 
and $\Delta v$ were then used for the fits to the
$^{13}$CO and CI spectra.  The results are given in Table 1 which
lists the velocity, linewidth, and the velocity
integrated line intensities ($I$) for each of the 
components (labeled A to E for reference).

\begin{deluxetable}{cccccccccc}
\footnotesize
\tablecaption{PROPERTIES OF THE OBSERVED GLOBULES \label{tbl-1}}
\tablewidth{0pt}
\tablehead{
\colhead{Globule} & \colhead{$v_o$} & \colhead{$\Delta v$} &
\colhead{$I$(CO)} & \colhead{$I$($^{13}$CO)} & \colhead{$I$(C\,I)} & 
\colhead{$I_{12}/I_{13}$} & \colhead{$N$(CO)} & 
\colhead{$N$(C\,I)} &
\colhead{$N$(C\,I)/$N$(CO)} \\
 & \multicolumn{2}{c}{\kms} & \multicolumn{3}{c}{\Kkms} & & 
   \multicolumn{2}{c}{10$^{15}$cm$^{-2}$} & 
}
\startdata
A & $-$21.9 & 1.7 & 3.4 & 0.26 & 0.93 & 13 & 1.80 & 12.2 & 6.8 \nl
B & $-$20.1 & 1.2 & 1.7 & 0.10 & 0.45 & 16 & 0.89 &  5.9 & 6.7 \nl
C & $-$17.3 & 2.4 & 2.4 & 0.17 & 0.63 & 14 & 1.27 &  8.3 & 6.5 \nl
D & $-$13.7 & 2.1 & 2.5 & 0.13 & 0.40 & 20 & 1.32 &  5.3 & 4.0 \nl
E & $-$11.1 & 1.4 & 1.6 & 0.16 & 0.44 & 11 & 0.87 &  5.8 & 6.7 \nl
\enddata
\end{deluxetable}

The CO mapping data are shown in Fig.~3 as a series of channel maps,
with a velocity spacing of 1~km~s$^{-1}$.
The CO emission is seen to extend along a south-east
north-west direction and shows a complex structure. The emission shows
peaks at different positions in different channel maps, indicating
that the distribution of matter is characterized by small globules or
clumps which have a typical velocity dispersion 
$\rm \lesssim \, 2 \, km \, s^{-1}$ 
and are unresolved or partially resolved by the 30\arcsec\ beam.  
Most of the components identified in the spectra
in Fig.~2 can be associated with discrete structures in the
channel maps.  Thus the bulk of the CO
and the C\,I emission is localized within the neutral globules.

\section{Discussion}
 
\subsection{C\,I and CO column densities}
We first discuss interpretation of the observed lines in terms of
column densities, using the simplifying assumptions of low optical
depth and LTE.
 The recent study of molecular gas in PN, including the
Helix, by Bachiller et al.\ (1997) indicates a relatively low opacity
for the CO lines and provides an estimate for the excitation
temperature of $T_{ex} = 25$ K, based on the ratio of the 2$-$1 and
1$-$0 lines of $^{13}$CO.  The low intensity of the C\,I line and the
similarity of the levels to the low lying CO lines suggests that these
results are also appropriate for C\,I.  For $T_{ex} = 25$~K, the C\,I
and CO column densities (in cm$^{-2}$) are related to the
integrated intensities of the observed lines (in {K\,\kms})
by the expressions $N({\rm C\,I}) = 1.3 \times 10^{16}\, I({\rm C\,I})$
and $N({\rm CO}) = 5.3 \times 10^{14}\, I({\rm CO})$. These relations
are not very sensitive to $T_{ex}$, and vary by
less than a factor of 2 over the range 10--80~K.

Estimated column densities are given in Table~1
together with the C\,I/CO column density ratio and the $I({\rm ^{12}
CO})/I({\rm ^{13} CO})$ ratio.  These results show that the
characteristics of the globules are remarkably uniform: the
$^{12}$C/$^{13}$C value is $\approx 15$, and the column density in
C\,I is systematically higher than that of CO by a factor of 
$\approx 6$.

\begin{figure}[h]
\vskip -0.95truecm
\plotone{figure3.ps}
\vskip -1truecm
\caption{Maps of the CO(2$-$1) line intensity integrated in velocity
intervals of 1~km s$^{-1}$. The central
lsr velocity of each interval is given in each
panel. The contours are  0.3, 0.6, 0.9 ... 3
K~km~s$^{-1}$. The bottom right panel
displays the velocity integrated emission 
(from $-$26 to $-$9~km~s$^{-1}$) - contours are 1 to 12 by 
3~K~km~s$^{-1}$. The beam (30$^{\prime\prime}$) shown in the 
upper left panel corresponds to the position observed in Fig.~2.
}\label{threebarrel}
\end{figure}

\subsection{Properties of the globules}

The large abundance of CI that we find in the Helix demonstrates that
atomic carbon is an important component of the envelope's neutral gas.
This is likely caused in part by the radiation field of the central
star which will develop PDRs in the neutral
gas.  The material observed lies along the ionization front in the
dark/extinguished area of the optical nebula (see Fig.~1) and the
morphology of the gas (Fig. 3) indicates a very broken structure which
is probably permeated by the stellar radiation.

The C/O ratio in the Helix has not been measured but there is
considerable evidence that it is carbon rich.
Howe et al. (1994) have constructed chemical models of the globules
and find that the abundance of C\,I is predicted to be roughly the same 
as that of CO for a large range in optical
depths if C/O$>$1. 
This is consistent with our observations.
For the case with C/O$<$1, however, the abundance of 
C\,I is predicted to be much less
than CO, which is not observed, indicating that the Helix is a C-rich 
nebula. Additional evidence comes from the similarity of the
molecular abundances in the Helix with other C-rich nebulae measured by
Bachiller et al. (1997), and the presence of PAH emission 
(Cox et al., in preparation).

The globules' sizes determined from the channel maps 
are typically $30\arcsec \times \lesssim 10 \arcsec$
or 7~$\times \lesssim 2$ in units of $10^{16}$~cm (at
160~pc). Their masses can be estimated from the
observed lines by adopting an abundance ratio for CO or C.
Estimates of the mass of {\it molecular gas} in the globules can be made 
assuming all the oxygen is in CO, i.e.,  CO/H = O/H which is measured to
be 3$\times$10$^{-4}$ in the ionized gas (Kaler et al. 1990). 
Using the same assumptions on the opacity and temperature as before, the beam
averaged column densities  of hydrogen (in H\,I and H$_2$) are  
$\approx 3\times$10$^{18}$~cm$^{-2}$ for clumps A, B, C, and E, and
the masses are  typically 1-3~$\times$10$^{-5}$~M$_{\odot}$
for each clump. We note that component D corresponds 
to material at the edge of a CO layer, which could explain
its lower N(C\,I)/N(CO) ratio. 
						 
The average value of the C\,I/CO ratio of $\approx 6$ 
that we find means that  
the above mass estimates based on the molecular observations are lower
limits to the total mass of {\it neutral gas}.
Adopting a nominal C/O ratio of 1.2 (as in the Ring nebula - see
Bachiller et al. 1994 and references therein), the corresponding
mass of neutral gas of each clump will be about 10$^{-4}$~M$_{\odot}$.
Taking the total number of clumps over the entire surface of the
nebula estimated to be 3000 by O'Dell \& Handron (1996) 
and assuming that they are similar to the clumps described in 
this paper, we derive
a total mass of neutral gas of 0.4~M$_{\odot}$ in the Helix.
This is an order of magnitude greater than the estimate published by Huggins
\& Healy (1986). This number could be overestimated because clumps inside
the ionized cavity are known to have smaller masses of 
typically 5$\times$10$^{-6}$~M$_{\odot}$ (Huggins et al. 1992). 
On the other hand, taking into account the contribution 
of the C\,II 158~$\rm \mu m$ fine structure line which is 
likely to be present in the neutral gas will increase the 
above mass estimate. Work in progress will explore the 
C\,II content in the Helix nebula. In any case, it is clear
that the mass of the neutral gas in the Helix represents a 
significant fraction, perhaps up to 50\%, of the total 
nebular mass. Comparable results were found for the Ring nebula
by Bachiller et al. (1994). which is in a similar evolutionary
status to the Helix.       
 
\subsection{Evolution}
The large abundance of C\,I in the Helix is very
different from that found at earlier 
stages of evolution. C\,I has not been detected in any envelopes of
AGB stars except IRC+10216,
which has only a small amount of CI in the outer envelope
(Keene et al. 1993).
The young PN NGC~7027 has a larger C\,I content (about
1/2 that of CO in the inner envelope) associated with the PDR
surrounding the compact ionized nebula (Young et al., in preparation).
This is consistent with the ionizing front gradually etching out 
the molecular envelope ejected during 
the AGB phase. Further evolution is seen in the Helix and
Ring (Bachiller et al. 1994) nebulae,
where vast amounts of C\,I are detected. 

Evidence that such an evolution has taken place inside the
Helix is also suggested by the difference in mass between the globules 
seen in the ionized cavity and those of the outer molecular envelope.
The mass of molecular gas derived for the globules in the envelope
are about a factor 2--6 times bigger than the mass estimates 
for the cometary globules which lie inside the ionized 
cavity of the Helix (Huggins et al. 1992). This difference 
can be understood in the general picture wherein the ionization 
front overtakes clumps already present in 
the molecular envelope and photoionizes their surface layers. 
As they are gradually etched
away, the dense molecular core will become smaller, and the 
remains, if any, will probably resemble the clumps detected 
in the ionized region. 
We note that this evolutionary picture is fully consistent with 
the results of the recent CO survey by Huggins et al. (1996).

\section{Conclusion}
These observations places important constraints on the
properties of the neutral gas in the Helix nebula. The
mapping observations show that the structure of the gas
on the periphery of the ionized nebula is highly fragmented,
leading naturally to the formation of cometary globules in
the ionized cavity. Our observations show that  C\,I 
is the major form of atomic carbon in the 
neutral gas of the Helix, and probably dominates the cooling of 
the gas. The C\,I results also indicate that the Helix is a 
carbon-rich nebula. The total mass of neutral gas in the 
Helix is derived to be of the order 0.4~M$_{\odot}$, an 
order of magnitude larger than previous estimates.   The present
findings are in agreement with current ideas on the evolution
from the AGB phase to fully evolved PN, during which the
neutral envelope ejected in the AGB phase is gradually
photodissociated and photoionized by the radiation field of
the hot central star.
 \\

We thank Dr. R. Martin of the RGO for providing the R-Band image.
This work was supported in part by NSF grant AST93-14408 
(to P.J.H.). The CSO is supported by NSF grant AST93--13929.

\end{document}